\documentclass[preprint,pra,nofootinbib]{revtex4}
\usepackage{mathrsfs}
\usepackage{graphicx}
\usepackage{epsfig}
\usepackage{dcolumn}
\usepackage{bm}
\usepackage{amsmath,amssymb,amsthm}
\usepackage[colorlinks=true,linkcolor=blue]{hyperref}
\usepackage{subfigure}
\usepackage{booktabs}
\usepackage[mathscr]{euscript}
\usepackage[normalem]{ulem}

\newcommand\redsout{\bgroup\markoverwith{\textcolor{red}{\rule[0.6ex]{6pt}{0.6pt}}}\ULon}

\begin{document}


\title{Peakons and pseudo-peakons of higher order b-family equations}
\author{Si-Yu Zhu$^1$, Ruo-Xia Yao$^{1*}$\thanks{Corresponding author}, De-Xing Kong$^2$ and S. Y. Lou$^{3,4*}$\thanks{Corresponding author: lousenyue@nbu.edu.cn}}
\address{\small{$^1$School of Computer Science, Shaansi Normal University, Xi'an, 710119, Shaanri, China\\
$^2$Zhejiang Qiushi Institute for Mathematical Medicine, Hangzhou 311121, China\\
$^3$School of Physical Science and Technology, Ningbo University, Ningbo, China \\
$^{4}$Institute of Fundamental Physics and Quantum Technology, Ningbo University, Ningbo, China}}



\date{\today}

\begin{abstract}
This paper explores the rich structure of peakon and pseudo-peakon solutions for a class of higher-order $b$-family equations, referred to as the $J$-th $b$-family ($J$-bF) equations. We propose several conjectures concerning the weak solutions of these equations, including a $b$-independent pseudo-peakon solution, a $b$-independent peakon solution, and a $b$-dependent peakon solution. These conjectures are analytically verified for $J \leq 14$ and/or $J \leq 9$ using the computer algebra software MAPLE. The $b$-independent pseudo-peakon solution is a 3rd-order pseudo-peakon for general arbitrary constants, with higher-order pseudo-peakons derived under specific parameter constraints. Additionally, we identify both $b$-independent and $b$-dependent peakon solutions, highlighting their distinct properties and the nuanced relationship between the parameters $b$ and $J$. The existence of these solutions underscores the rich dynamical structure of the $J$-bF equations and generalizes previous results for lower-order equations. Future research directions include higher-order generalizations, rigorous proofs of the conjectures, interactions between different types of peakons and pseudo-peakons, stability analysis, and potential physical applications. These advancements significantly contribute to the understanding of peakon systems and their broader implications in mathematics and physics.
\end{abstract}


\maketitle
\section{Introduction}
The b-family of equations represents a significant class of nonlinear partial differential equations that have garnered substantial attention in the field of mathematical physics. These equations are known for their rich structure and the presence of peakon solutions, which are solitary waves with a peak at their crest. The general form of the b-family equation is given by:
\begin{equation}
m_t+vm_x+bv_xm=0,\quad m=(1-\partial_{x})^2v,
\end{equation}
where $b$ is a parameter that influences the nonlinearity and dispersion properties of the equation.
The study of the b-family equations has provided significant insights into nonlinear wave behavior, particularly through the discovery of peakon solutions (weak solutions) with discontinuous first derivatives at their peaks. These solutions have been widely studied due to their unique properties and physical applications.

Camassa and Holm \cite{CH} introduced the Camassa-Holm equation, a special case of the b-family with $b=2$, which admits peakon solutions. This foundational work spurred further research into the integrability and stability of peakons. Degasperis and Procesi \cite{DP} derived another member of the b-family for $b=3$, the Degasperis-Procesi equation, which also supports peakons and exhibits a bi-Hamiltonian structure, emphasizing its integrability.

The concept of pseudo-peakons, smooth approximations of peakons, has further enriched the field. Research by Lenells \cite{Lenells} and others has explored their properties and stability, bridging the gap between smooth solitons and peakons. Constantin and Strauss \cite{CS} investigated peakon stability, while Ivanov \cite{Ivanov} advanced the understanding of higher-order b-family equations. Holm and Staley \cite{HS} analyzed wave structures and nonlinear balances in evolutionary PDEs, contributing to the dynamics of peakon systems. Johnson \cite{Johnson} connected the Camassa-Holm and Korteweg-de Vries equations, enhancing their relevance in water wave modeling. Olver-Rosenau \cite{OR} and Chen-Liu-Zhang \cite{CLZ} generalized the Camassa-Holm equation to two components, expanding solution scopes and integrability insights.

Recent advancements have extended the b-family and related models to higher-order formulations \cite{HIS,Lou,Fokas}. Henry, Ivanov, and Sakellaris \cite{HIS} extended the CH and DP equations, offering a theoretical framework for oceanic internal wave-current interactions. Gorka, Pons, and Reyes \cite{GPR} explored higher-order Camassa-Holm equations using loop group geometry, while Liu and Qiao \cite{LiuQiao} introduced pseudo-peakons and multi-peakons in a fifth-order model. Qiao and Reyes \cite{QR} further advanced the field with fifth-order extensions, broadening the understanding of nonlinear wave phenomena.

In this paper, we investigate the peakon and pseudo-peakon structures for a class of higher-order b-family equations, defined as
\begin{equation}
m_t + v m_x + b v_x m = 0, \quad m = (1 - \partial_{x}^2)^J v,
\label{JbF}
\end{equation}
where $J$ is an arbitrary positive integer. For notational convenience, we refer to \eqref{JbF} as the $J$-th b-family ($J$-bF) equation. Specifically, when $b = 2$, it is termed the $J$-th Camassa-Holm ($J$-CH) equation, and for $b=3$, it is referred to as the $J$-th Degasperis-Procesi ($J$-DP) equation. The loop group geometric aspects of the $J$-CH equation have been previously explored in \cite{GPR} for more general high order peakon systems with $m=\sum_{i=0}^nc_i\partial_x^{2i}v$.

In Section II of this paper, we propose several conjectures concerning the weak solutions of the $J$-bF system \eqref{JbF}. Sections III through VI are dedicated to verifying the validity of these conjectures and the conclusions presented in Section II for the cases $J = 2$, $3$, $4$, and $5$, respectively. The final section provides a brief summary and additional discussions.

\section{Conjectures}
Through extensive and intricate calculations on the weak solutions of the $J$-bF system \eqref{JbF} for various values of $J$, we propose the following conjectures:\\
\bf Conjecture 1. \rm \textit{Pseudo-peakon conjecture.}
The $J$-bF equation \eqref{JbF} admits a $b$-independent pseudo-peakon solution
\begin{equation}
v = c \left(1 + |\xi| + \sum_{i=1}^{J-2} a_i |\xi|^{i+1} \right) \mbox{\rm e}^{-|\xi|}, \quad \forall J \geq 3,\ \quad \xi\equiv x-ct,
\label{Conj}
\end{equation}
where $a_1, a_2, \dots, a_{J-2}$ are arbitrary constants.

For $J=2$, the pseudo-peakon solution can also be cast into the conjecture without the summation term. The conjecture has been verified for the cases of $J\leq 14$ by means of the computer algebra MAPLE. While a clear path to a general proof remains elusive at this stage, we are hopeful that a proof will be established by researchers in the near future.

The simplest scenario, corresponding to \( J= 2 \) in equation \eqref{Conj}, has been previously established by several researchers. However, the generalized solution \eqref{Conj} for any \( J > 2 \) is introduced for the first time in this paper. The pseudo-peakon solution \eqref{Conj} exhibits a rich and intricate structure. In this paper, we explore specific instances for various values of \( J \). Prior to delving into particular cases, we provide a formal definition of the \(n\)th-order pseudo-peakon.\\
\bf Definition. \rm A solution is termed the \( n \)th-order pseudo-peakon if its \( i \)th-order \( x \)-derivatives are continuous for all \( i \leq n - 1 \), while the \( n \)th-order \( x \)-derivative is discontinuous.

By examining the smoothness of the solution \eqref{Conj}, we can identify several properties of the pseudo-peakon \eqref{Conj}:
\begin{enumerate}
    \item[(i)] The solution \eqref{Conj} represents a $3$rd-order pseudo-peakon for general values of $a_i$ ($i = 1, 2, \ldots, J-2$), except in the following special cases.

    \item[(ii)] If the constants $a_1$ and $a_2$ satisfy the constraint condition ($v_{xi}=v_{x^i}$), $\left.v_{x3}\right|_{\xi\rightarrow 0^+}=\left.v_{x3}\right|_{\xi\rightarrow 0^-}$, i.e.,
    \begin{equation}
        1 - 3(a_1 - a_2) = 0, \label{a12}
    \end{equation}
    the solution \eqref{Conj} becomes a $5$th-order pseudo-peakon without requiring additional constraints on the remaining parameters.

    \item[(iii)] A $7$th-order pseudo-peakons can be derived from \eqref{Conj} by imposing the parameter constraints \eqref{a12} and
    \begin{equation}
        1 - 5a_1 + 15a_2 - 30a_3 + 30a_4 = 0, \label{a34}
    \end{equation}
    which is obtained from $\left.v_{x5}\right|_{\xi\rightarrow 0^+}=\left.v_{x5}\right|_{\xi\rightarrow 0^-}$.

    \item[(iv)] Under the parameter constraints \eqref{a12}, \eqref{a34}, and
    \begin{equation}
        1 - 7a_1 + 35a_2 - 140a_3 + 420a_4 - 840a_5 + 840a_6 = 0, \label{a56}
    \end{equation}
    which is equivalent to $\left.v_{x7}\right|_{\xi\rightarrow 0^+}=\left.v_{x7}\right|_{\xi\rightarrow 0^-}$,
    $9$th-order pseudo-peakons can be obtained from \eqref{Conj}.
\end{enumerate}

Higher-order pseudo-peakons can be derived from \eqref{Conj} by introducing additional parameter constraints which can be derived from
\begin{equation}
\left.v_{x^{2j+1}}\right|_{\xi\rightarrow 0^+}=\left.v_{x^{2j+1}}\right|_{\xi\rightarrow 0^-}.\label{constraint}
\end{equation}
For odd $J=2n+1$ (even $J=2(n+1)$), one can obtain  3rd, 5th, ..., $(4n+1)$-th ($(4n+3)$-th) order pseudo-peakon solutions by suitable parameter conditions from \eqref{constraint}.

In addition to the pseudo-peakon solution \eqref{Conj}, we may identify various other types of weak solutions. Here, we present two conjectures on peakons in this section:\\
\textbf{Conjecture 2.} \textit{($b$-independent peakon conjecture).}
The $J$-bF equation \eqref{JbF} admits a $b$-independent peakon solution of the form
\begin{equation}
    v = c \left(1 + \sum_{i=1}^{J-1} a_i |x - ct|^{i} \right) \mathrm{e}^{-|x - ct|}, \quad \forall J \geq 1,
    \label{Conj2}
\end{equation}
where the $b$-independent constants $a_i$ satisfy the inequalities $0 < a_{J-1} < a_{J-2} < \cdots < a_2 < a_1 < 1$.

Conjecture 2 has been verified for $J \leq 9$. However, a unified formula for all $J$ remains elusive. The first few sets of constants $a_i$ are given by
\begin{eqnarray}
    &&\{J=2: \ a_1=2\},\ \left\{J=3: \ a_1=\frac{72}{138},\ a_2=\frac{13}{138}\right\},\nonumber\\
    &&\left\{J=4: \ a_1=\frac{495}{1136},\ a_2=\frac{81}{1136},\ a_3=\frac{1}{213}\right\},\nonumber\\
    &&\left\{J=5: \ a_1=\frac{16433970}{29110637},\ a_2=\frac{40909737}{291106370},\ a_3=\frac{8220074}{436659555},\ a_4=\frac{274669}{232885096}\right\}.
    \label{aa}
\end{eqnarray}
\textbf{Conjecture 3.} \textit{($b$-dependent peakon conjecture).}
The $J$-bF equation \eqref{JbF} admits one real $b$-dependent peakon solution for odd $J$ and two real $b$-dependent peakon solutions for even $J$, expressed in the form
\begin{equation}
    v = c \left(\sum_{i=0}^{J-1} c_i |x - ct|^{i} \right) \mathrm{e}^{-|x - ct|}, \quad \forall J \geq 3,
    \label{Conj3}
\end{equation}
where the $b$-dependent constants $c_i$ ($i = 0, 1, 2, \ldots, J-1$) are determined for each $J$, with $c_0 \neq 1$.

The validity of this conjecture has been verified for $J \leq 9$. For $J = 3$, the constants $c_i$ are given by
\begin{eqnarray}
    c_0 &=& \frac{385923}{385923 - 67195b}, \quad c_1 = \frac{188717}{385923 - 67195b}, \quad c_2 = \frac{30038}{385923 - 67195b}.
    \label{ccJ3}
\end{eqnarray}
For $J = 4$, the constants $c_i$ in \eqref{Conj3} are
\begin{eqnarray}
    c_1 &=& \left(\frac{317039585}{727147048} + \frac{9089301767}{17451529152b}\right)c_0 - \frac{9089301767}{17451529152b}, \nonumber \\
    c_2 &=& \left(\frac{78194177}{1090720572} + \frac{813796249}{3272161716b}\right)c_0 - \frac{813796249}{3272161716b}, \nonumber \\
    c_3 &=& \left(\frac{7045177}{1454294096} + \frac{3931438217}{104709174912b}\right)c_0 - \frac{3931438217}{104709174912b},
    \label{ccJ4}
\end{eqnarray}
where $c_0$ is determined by the quadratic equation
\begin{equation*}
    \frac{19767871994017357}{55512146191680}(c_0 - 1)^2 + \frac{17680946891729}{385501015220}bc_0(c_0 - 1) + b^2c_0^2 = 0,
\end{equation*}
which yields two real roots for arbitrary $b$.

Through extensive calculations for larger $J$ and fixed real $b$, it is observed that there exists exactly one (two) real $b$-dependent peakon solution(s) \eqref{Conj3} for odd $J = 2n + 1$ (even $J = 2n + 2$), alongside $2(n - 1)$ complex peakons. However, the complex peakons are not discussed in this paper.

\section{Fifth order \(b\) family and its peakon and pseudo-peakon solutions}
For $J = 2$, the $J$-bF system \eqref{JbF} becomes one of the known fifth order $b$-families ($v_{xi}=v_{x^i}$)
\begin{equation}
m_t + v m_x + b v_x m = 0, \quad m = v-2v_{xx}+v_{x4}.
\label{JbF2}
\end{equation}
The single weak peakon and/or the pseudo-peakon solution of \eqref{JbF2} can be assumed in the form
\begin{equation}
v = c(a_0 +a_1 |\xi|) \mbox{\rm e}^{-|\xi|},\quad \xi=x-ct, \label{J2pp}
\end{equation}
which is hinted by vanishing $m$.

From \eqref{J2pp}, we have
\begin{eqnarray}
v_x&=&-c\ \mathrm{sgn}(\xi)(a_1|\xi|+a_0-a_1)\mathrm{e}^{-|\xi|},\ \nonumber\\
m&=& 2\delta(\xi)\{(a_1|\xi|+a_0-2a_1)[6\delta(\xi)-4\ \mathrm{sgn}(\xi)]+5a_1|\xi|+5a_0-17a_1\}
\mathrm{e}^{-|\xi|},\nonumber\\
m_t&=&-cm_x=-c\left\{8[\mathrm{sgn}(\xi)\delta'(\xi)+2\delta(\xi)^2
-3\delta(\xi)\delta'(\xi)](2a_1-a_1|\xi|-a_0)\right.\nonumber\\
&&
-4\delta(\xi)[3\ \mathrm{sgn}(\xi)\delta(\xi)-2](a_1|\xi|+a_0-3a_1)
-2(5a_1|\xi|+5a_0-22a_1)\ \mathrm{sgn}(\xi)\delta(\xi)\nonumber\\
&&\left.
+2(5a_1|\xi|+5a_0-17a_1)\delta(\xi)\right\}\mathrm{e}^{-|\xi|},\label{xi}
\end{eqnarray}
where $\mathrm{sgn}(\xi)$ is the signum function of $\xi$, $\delta(\xi)$ is the Dirac delta function of $\xi$ while $\delta'(\xi)$ is the derivative of $\delta(\xi)$ with respect to $\xi$.

Substituting \eqref{J2pp} into \eqref{JbF2} and requiring the result is a zero distribution yields only two possible solutions
\begin{equation}
a_0=a_1=1 \label{a011}
\end{equation}
and
\begin{equation}
a_0=1,\ a_1=\frac12. \label{a012}
\end{equation}
The first case is just the known pseudo-peakon solution
\begin{equation}
v = c(1 + |\xi|) \mbox{\rm e}^{-|\xi|},\quad \xi=x-ct, \label{J2pp1}
\end{equation}
which shares the same functional form as a solution to another fifth-order CH equation \cite{QR}. The solution \eqref{J2pp1} is classified as a 3rd pseudo-peakon but not a peakon (1st pseudo-peakon). This distinction arises because the first and second order derivatives of $u$ are continuous, satisfying
$v_x(\xi = 0^+) = v_x(\xi = 0^-) \quad \text{and} \quad v_{xx}(\xi = 0^+) = v_{xx}(\xi = 0^-),$
while the third order derivative is discontinuous, i.e.,
$v_{x3}(\xi = 0^+) \neq v_{x3}(\xi = 0^-).$

The second solution \eqref{a012} yields a previously unknown $b$-independent peakon solution
\begin{equation}
v = \frac{1}{2}c\left(2 + |\xi|\right) \mbox{\rm e}^{-|\xi|}, \label{J2p}
\end{equation}
for the $2$-bF equation \eqref{JbF2}.
\section{Seventh order \(b\) family and 3rd and 5th order pseudo-peakons}
For $J=3$, the model \eqref{JbF} becomes a seventh order $b$-family
\begin{eqnarray}
&&m_t + v m_x + b v_x m = 0, \quad m = v - 3v_{xx} + 3v_{x4} - v_{x6}. \label{3bF}
\end{eqnarray}
By vanishing $m$ of \eqref{3bF}, one can assume that the seventh order $b$-family equation \eqref{3bF} admits
the weak solutions in the form
\begin{eqnarray}
&&v=c\left(a_0 + a_1|x - ct| + a_2|x - ct|^2\right)\mbox{\rm e}^{-|x - ct|}, \label{J3pp}
\end{eqnarray}
for possible parameter selections.

Substituting \eqref{J3pp} into the seventh order $b$-family equation \eqref{3bF}, one can find that requiring the result is a zero distribution only for the following conditions
\begin{eqnarray}
&&(a_0-1)(a_0-3a_1+6a_2)=0,\nonumber\\
&&15(a_0-a_1)(a_0-3a_1+6a_2)b+(a_0-1)(59a_0-223a_1+643a_2)=0,\nonumber\\
&&2(a_0-a_1)(7a_0-34a_1+114a_2)b+(a_0-1)(27a_0-149a_1+646a_2)=0.\label{a012J3}
\end{eqnarray}
From \eqref{a012J3}, it is not difficult to find three sets of solutions,
\begin{eqnarray}
&&a_0=a_1=1, \label{j31}\\
&&a_0=1,\ a_1=\frac{12}{23},\ a_2=\frac{13}{138},\label{j32}\\
&&a_0=\frac{385923}{385923-67195b},\ a_1 = \frac{188717}{385923-67195b},\ a_2=\frac{30038}{385923-67195b}.\label{j33}
\end{eqnarray}

By substituting solution \eqref{j31} back into \eqref{J3pp}, we obtain the result of the pseudo-peakon that coincides with the conjecture \eqref{Conj}, expressed as:
\begin{eqnarray}
&&v = c\left(1 + |x - ct| + a_1|x - ct|^2\right)\mbox{\rm e}^{-|x - ct|}, \label{J3pp1}
\end{eqnarray}
where the arbitrary constant $a_2$ has been redefined as $a_1$. This formulation captures the essence of the pseudo-peakon solution, emphasizing its dependence on the arbitrary parameter $a_1$.

The solution \eqref{J3pp1} represents a 3rd-order pseudo-peakon for a general constant $a_1 \neq \frac{1}{3}$, while it reduces to a 5th order pseudo-peakon for the specific constant $a_1=\frac{1}{3}$. Figs. 1(a)--1(d) illustrate the intricate structure of the pseudo-peakon \eqref{J3pp1} for parameter choices $a=-2$, $-50$, $2$, and $\frac{1}{2}$, respectively, at time $t = 0$. The velocity parameter $c$ and the time $t$ are fixed at $c=1$ and $t=0$ for all figures in this paper. Fig. 1(e) demonstrates the structure of the second-order partial derivative, $v_{xx}$, at the center of the pseudo-peakon \eqref{J3pp1} with $a=\frac{1}{2}$. Similarly, Fig. 1(f) highlights the property of the fourth-order partial derivative, $v_{x4}$, at the center of the pseudo-peakon \eqref{J3pp1} with $a=\frac{1}{3}$. The Figures 1(e) and 1(f) indicate also the discontinuous of the third and the fifth order partial derivatives of the field $v$ with respect to $x$.

\begin{figure}
\begin{center}
\subfigure{
\includegraphics[width=0.45\textwidth,height=3cm]{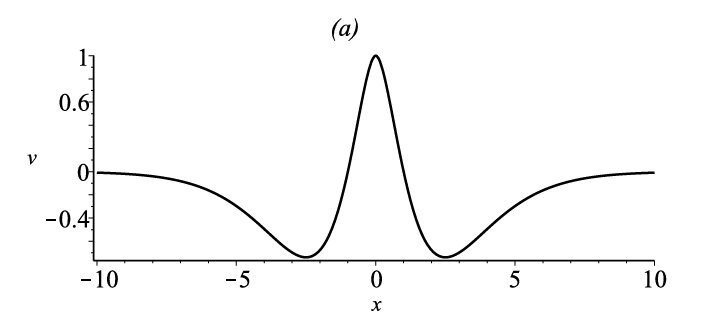}}
\centering \subfigure{
\includegraphics[width=0.45\textwidth,height=3cm]{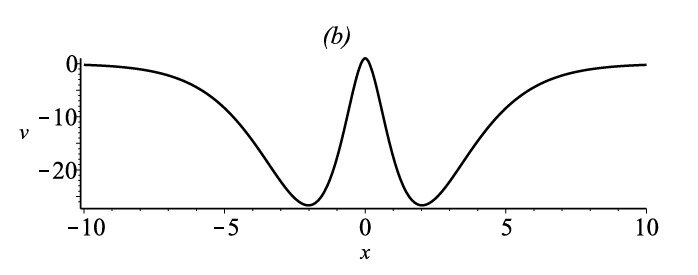}}
\centering\subfigure{
\includegraphics[width=0.45\textwidth,height=3cm]{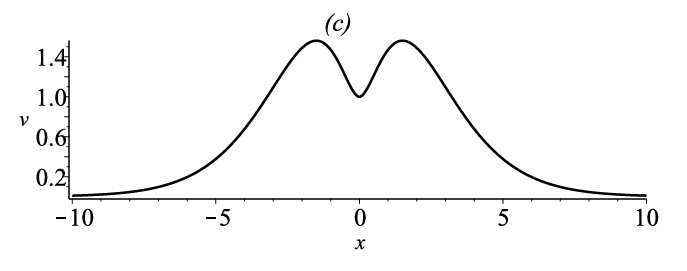}}
\centering \subfigure{
\includegraphics[width=0.45\textwidth,height=3cm]{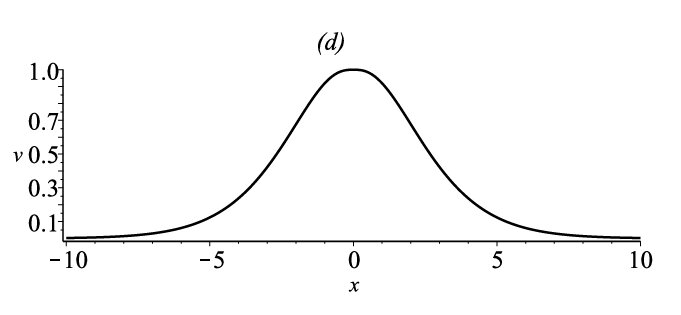}}
\centering \subfigure{
\includegraphics[width=0.45\textwidth,height=3cm]{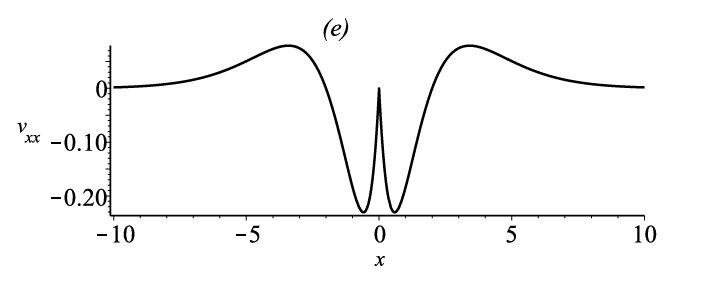}}
\centering \subfigure{
\includegraphics[width=0.45\textwidth,height=3cm]{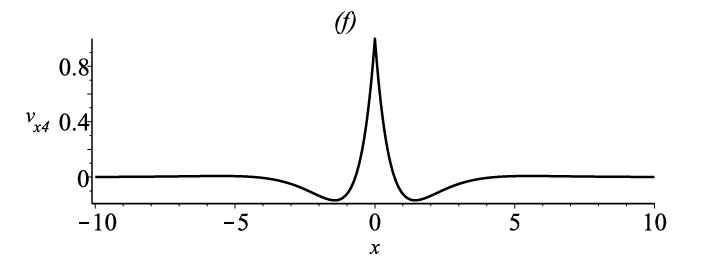}}
\caption{
    (a) Pseudo-peakon structure with one smooth peak and double bottoms expressed by \eqref{J3pp} and $a = -2$;
    (b) W-shaped pseudo-peakon structure given by \eqref{J3pp1} and $a = -50$;
    (c) M-shaped pseudo-peakon structure of \eqref{J3pp1} with $a = 2$;
    (d) Pseudo-peakon structure with a single smooth peak expressed by \eqref{J3pp1} and $a = \frac{1}{2}$;
    (e) The structure of the second-order derivative of the 3rd-order pseudo-peakon \eqref{J3pp1} with $a = \frac{1}{2}$;
    (f) The structure of the fourth-order derivative of the 5th-order pseudo-peakon \eqref{J3pp1} with $a = \frac{1}{3}$.
}
\label{fig1}
\end{center}
\end{figure}

By substituting the solution \eqref{j32} into the solution ansatz \eqref{J3pp}, we derive the result of the $b$-independent peakon conjecture (Conjecture 2 of Section II), which takes the following form:
\begin{equation}
v = \frac{1}{138}c\left(138 + 72|x - ct| + 13|x - ct|^2\right) \mathrm{e}^{-|x - ct|}. \label{J3p}
\end{equation}
This expression represents the peakon solution independent of the parameter $b$.
The structure of this \( b \)-independent peakon \eqref{J3p} is illustrated in Fig.~2(a) for the amplitude \( c = 1 \) at the time \( t = 0 \).

By substituting the solution \eqref{j33} into the solution assumption \eqref{J3pp}, we arrive at the result of the $b$-dependent peakon conjecture (Conjecture 3 of Section II), which is expressed as follows:
\begin{equation}
v = \frac{653c}{385923 - 67196b}\left(591 + 289|x - ct| + 46|x - ct|^2\right) \mathrm{e}^{-|x - ct|}. \label{J3bp}
\end{equation}
This equation describes the peakon solution, highlighting its dependence on the model parameter $b$

Fig.~2(b) depicts the structure of this \( b \)-dependent peakon \eqref{J3bp} with the amplitude \( \frac{385923c}{385923 - 67196b},\ b=3,\ c=1 \) at time \( t = 0 \). For this solution, a critical value of \( b \) exists, defined as \( b_{\text{cr}} = \frac{385923c}{67196} \approx 5.743329c \). When the parameter \( b \) transitions from \( b > b_{\text{cr}} \) (with \( c > 0 \)) to \( b < b_{\text{cr}} \), the peakon transforms into an anti-peakon. Moreover, as \( b \) approaches the critical value \( b_{\text{cr}} \), the amplitude of the peakon diverges to infinity.
\begin{figure}
\begin{center}
\subfigure{
\includegraphics[width=0.45\textwidth,height=3cm]{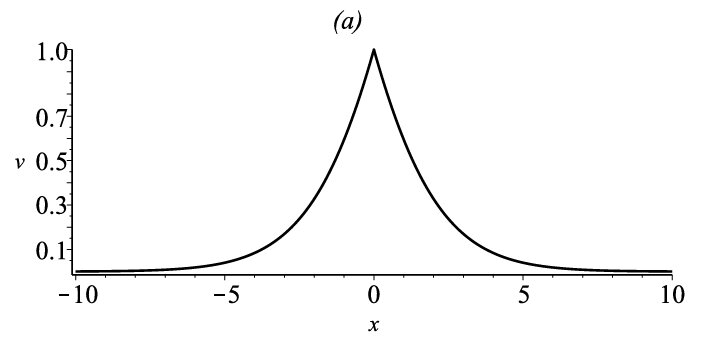}}
\centering \subfigure{
\includegraphics[width=0.45\textwidth,height=3cm]{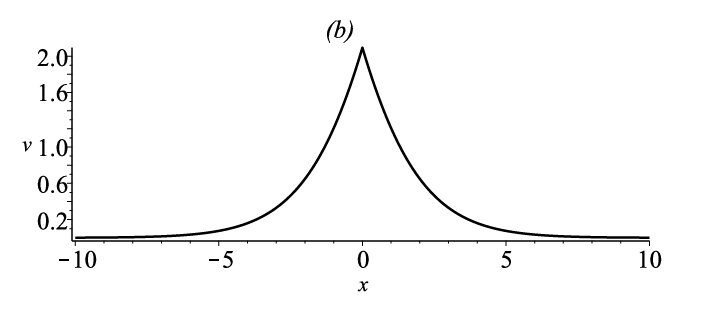}}
\caption{The graphs of peakons. 
    (a) The structure of the $b$-independent peakon expressed by \eqref{J3p};
    (b) The structure of the $b$-dependent peakon of \eqref{J3bp} with $b = 3$.
}
\label{fig2}
 \end{center}
\end{figure}
\section{Ninth order \(b\) family and $3$rd, $5$th, and $7$th order pseudo-peakons}
For $J=4$, the $J$-bF system \eqref{JbF} reduces to a ninth-order
$b$-family equation,
\begin{align}
m_t + v m_x + b v_x m = 0, \quad m = v - 4v_{xx} + 6v_{x4} - 4v_{x6} + v_{x8}, \label{4bF}
\end{align}
which admits the weak solution ansatz
\begin{align}
v = c\left(a_0 + a_1 |x - ct| + a_2|x - ct|^2 + a_3|x - ct|^3\right)\mathrm{e}^{-|x - ct|}. \label{J4PP}
\end{align}
Substituting \eqref{J4PP} into the ninth order
$b$-family equation \eqref{4bF}, the zero distribution condition yields
\begin{eqnarray}
&&60B(a_0-4a_1+12a_2-24a_3)b + A(377a_0-1671a_1+6042a_2-16842a_3) = 0, \nonumber\\
&&16B(24a_0-107a_1+389a_2-1089a_3)b + A(1259a_0-6214a_1+26038a_2-89712a_3) = 0, \nonumber\\
&&8B(5a_0-32a_1+176a_2-768a_3)b + A(81a_0-553a_1+3320a_2-16872a_3) = 0, \nonumber\\
&&A(a_0-4a_1+12a_2-24a_3) = 0, \quad A \equiv a_0-1, \quad B \equiv a_0-a_1. \label{J4aa}
\end{eqnarray}
It is evident that the simplest solution of \eqref{J4aa}, $A=B=0$ (i.e., $a_0=a_1=1$), corresponds to the pseudo-peakon solution,
\begin{align}
v = c\left(1 + |x - ct| + a_1|x - ct|^2 + a_2|x - ct|^3\right)\mathrm{e}^{-|x - ct|}, \label{J4pp}
\end{align}
which aligns with the solution presented in Conjecture 1 \eqref{Conj}. Here, the arbitrary constants
$a_3$ and $a_4$ have been redefined as $a_1$ and $a_2$, respectively.

The solution \eqref{J4pp} corresponds to a 3rd-order pseudo-peakon for general constants $\{a_1, a_2, b, c\}$ under the condition $3(a_2 - a_1) + 1 \neq 0$. When $3(a_2 - a_1) + 1 = 0$ but $a_1 \neq \frac{2}{5}$, the solution \eqref{J4pp} reduces to a 5th-order pseudo-peakon. Additionally, for the specific parameter values $\{a_1 = \frac{2}{5}, a_2 = \frac{1}{15}\}$, \eqref{J4pp} describes a 7th-order pseudo-peakon.

Figures 3(a)--3(d) illustrate the structures of the pseudo-peakon defined by \eqref{J4pp} for the parameter sets $\{a_1 = \frac{1}{2}, a_2 = \frac{1}{6}\}$, $\{a_1 = 1, a_2 = 1\}$, $\{a_1 = -\frac{5}{4}, a_2 = 1\}$, and $\{a_1 = 1, a_2 = -30\}$, respectively. To demonstrate the discontinuous nature of the pseudo-peakon \eqref{J4pp}, Figures 3(e)--3(f) showcase the peak characteristics of the fourth and sixth order derivatives of the field $v$ with respect to $x$.

\begin{figure}
\begin{center}
\subfigure{
\includegraphics[width=0.45\textwidth,height=3cm]{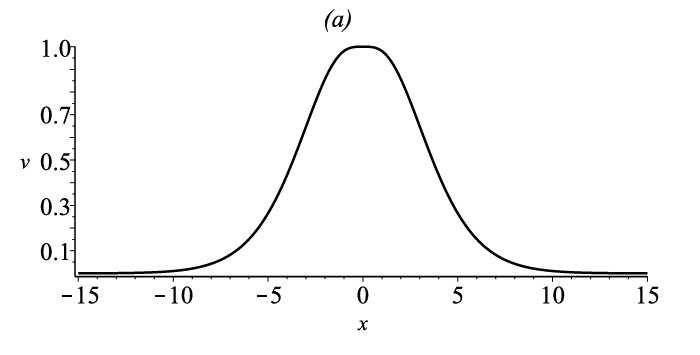}}
\centering \subfigure{
\includegraphics[width=0.45\textwidth,height=3cm]{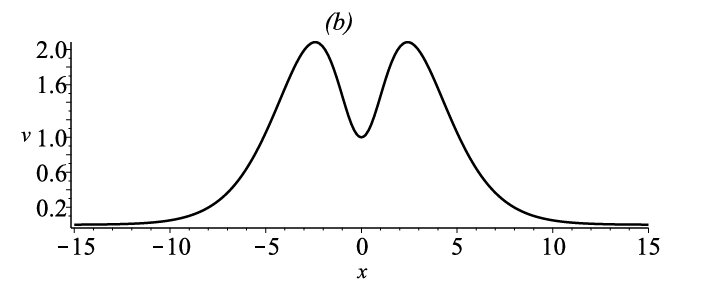}}
\subfigure{
\includegraphics[width=0.45\textwidth,height=3cm]{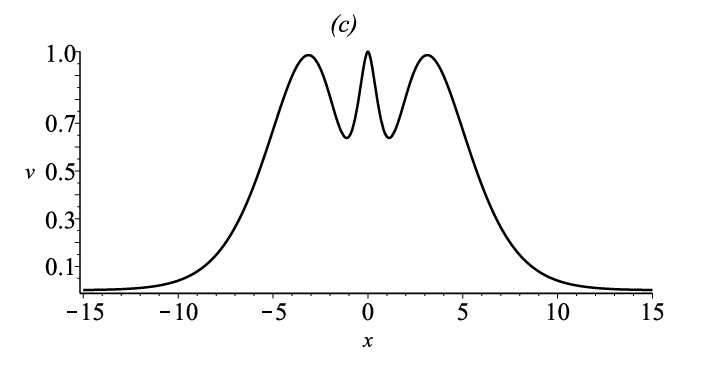}}
\centering \subfigure{
\includegraphics[width=0.45\textwidth,height=3cm]{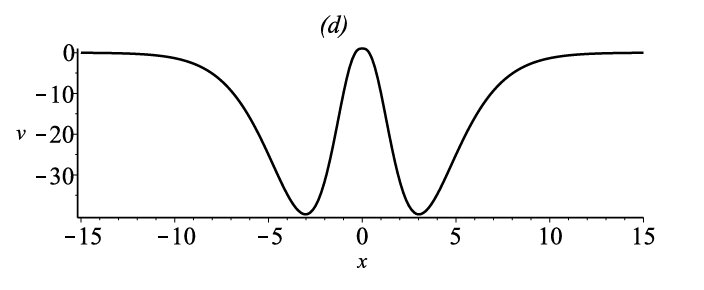}}
\subfigure{
\includegraphics[width=0.45\textwidth,height=3cm]{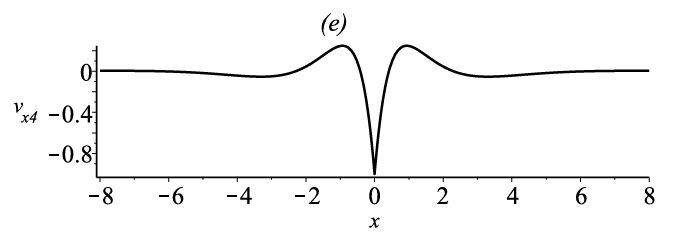}}
\centering \subfigure{
\includegraphics[width=0.45\textwidth,height=3cm]{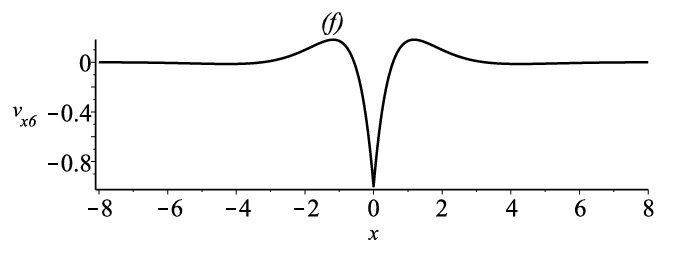}}
\caption{
    (a) A quite smooth pseudo-peakon \eqref{J4pp} with $\{a_1 = \frac{1}{2},\ a_2 = \frac{1}{6}\}$ and $v_x|_{x=ct} = v_{xx}|_{x=ct} = v_{xxx}|_{x=ct} = 0$;
    (b) M-shaped pseudo-peakon \eqref{J4pp} with $a_1 = a_2 = 1$;
    (c) The structure of the pseudo-peakon with three smooth peaks expressed by \eqref{J4pp}, $a_1 = -\frac{5}{4}$ and $a_2 = 1$;
    (d) W-shaped pseudo-peakon \eqref{J4pp} with $\{a_1 = 1,\ a_2 = -30\}$;
    (e) The structure of the fourth order derivative of $v$ with respect to $x$ for the 5th-order pseudo-peakon expressed by \eqref{J4pp} with the same parameters as (a);
    (f) The structure of the sixth order derivative of $v$ with respect to $x$ for the 7th-order pseudo-peakon given by \eqref{J4pp} with the selections $\{a_1 = \frac{2}{5},\ a_2 = \frac{1}{15}\}$.
}
\label{fig3}
 \end{center}
\end{figure}

The second solution of \eqref{J4aa} corresponds to $A=0$ and $B\neq 0$, which yields the $b$-independent solution \eqref{ccJ3}. This solution represents the $b$-independent peakon solution, expressed as:
\begin{align}
v = \frac{c}{3408}\left(3408 + 1485 |x - ct| + 243|x - ct|^2 + 16|x - ct|^3\right)\mathrm{e}^{-|x - ct|}. \label{J4p}
\end{align}

The third solution of \eqref{J4aa} corresponds to $A\neq0$ and $B\neq 0$, which yields the
$b$-dependent peakon solution \eqref{ccJ4}. This solution is associated with two
$b$-dependent peakon solutions, and it can be expressed in a unified form (with $\alpha\equiv 11461919 (a_0-1))$ as:
\begin{eqnarray}
v &=& \frac{c}{104709174912b}\left[24a_0b\left(21135531|x-ct|^3 + 312776708|x-ct|^2 + 1902237510|x-ct|\right.\right.\nonumber\\
&&\left.\left. + 4362882288\right) + \alpha\left(343|x-ct|^3 + 2272|x-ct|^2 + 4758|x-ct|\right)\right]\mathrm{e}^{-|x - ct|}, \label{J4pb}
\end{eqnarray}
where the parameter $a_0$  takes two distinct values:
\begin{align}
a_0 = \frac{1}{27905650366969405}\left(23586956522630821 \pm 288\sqrt{56775579142818640974199474}\right). \label{a0pm}
\end{align}
The structures of the solutions \eqref{J4p} and \eqref{J4pb} are qualitatively similar to those illustrated in Figs. 2(a)--2(b).
\section{Eleventh order \(b\)-family and 3rd, 5th, 7th, and 9th order pseudo-peakons}
For \( J = 5 \), the \( J \)-bF system \eqref{JbF} reduces to an eleventh order \( b \)-family equation:
\begin{align}
    m_t + v m_x + b v_x m = 0, \quad m = v - 5v_{xx} + 10v_{x4} - 10v_{x6} + 5v_{x8} - v_{x10}, \label{5bF}
\end{align}
which admits the following weak solution ansatz:
\begin{align}
    v = c\left(a_0 + a_1 |x - ct| + a_2|x - ct|^2 + a_3|x - ct|^3 + a_4|x - ct|^4\right)\mathrm{e}^{-|x - ct|}. \label{J5PP}
\end{align}
Here, the constants \( a_i \) (\( i = 0, 1, \ldots, 4 \)) are to be determined by requiring that the solution \eqref{J5PP} satisfies the weak solution condition of \eqref{5bF}.

Substituting \eqref{J5PP} into the eleventh order \( b \)-family equation \eqref{5bF} and imposing the condition that the resulting equation vanishes in the distributional sense, we derive the following constraints:
\begin{eqnarray}
&&90 B C b +A (857 a_0-4418 a_1+19194 a_2-68088 a_3+186888 a_4)=0,\nonumber\\
&&2 B (61 a_0-466 a_1+3170 a_2-18960 a_3+96240 a_4) b\nonumber\\
&&\quad +A (243 a_0-1985 a_1+14554 a_2-94890 a_3+540240 a_4)=0,\nonumber\\
&&5 B (363 a_0-1985 a_1+9458 a_2-38466 a_3+129360 a_4) b\nonumber\\
&&\quad +A (5648 a_0-33181 a_1+172801 a_2-786780 a_3+3069900 a_4) =0,\nonumber\\
&&28 B (197 a_0-968 a_1+3984 a_2-13368 a_3+34968 a_4) b\nonumber\\
&&\quad +A (27127 a_0-143737 a_1+654694 a_2-2511654 a_3+7899360 a_4)=0,
\nonumber\\
&&AC=0,\ C\equiv a_0-5 a_1+20 a_2-60 a_3+120 a_4, \label{aJ5}
\end{eqnarray}
where $A$ and $B$ are same as in \eqref{J4aa}.

From the constraint equation \eqref{aJ5}, it is evident that the simplest solution \( A = B = 0 \) leads to the result of Conjecture 1:
\begin{align}
    v = c\left(1 + |x - ct| + a_1|x - ct|^2 + a_2|x - ct|^3 + a_3|x - ct|^4\right)\mathrm{e}^{-|x - ct|}, \label{J5pp}
\end{align}
where the renamed constants \( a_1 \), \( a_2 \), and \( a_3 \) are arbitrary.

The solution \eqref{J5pp} represents a $b$-independent 3rd-order pseudo-peakon for arbitrary constants $\{a_1, a_2, a_3, b, c\}$, provided that $3(a_2 - a_1) + 1 \neq 0$. When $3(a_2 - a_1) + 1 = 0$ but $5(3a_3 - a_1) + 2c \neq 0$, the solution \eqref{J5pp} simplifies to a 5th-order pseudo-peakon. For $\{3(a_2 - a_1) + 1 = 0, 5(3a_3 - a_1) + 2c = 0\}$ but with $7a_1 \neq 3c$, the solution \eqref{J5pp} corresponds to a 7th-order pseudo-peakon. Finally, for the specific parameter values $\{a_1 = \frac{3}{7}, a_2 = \frac{2}{21}, a_3 = \frac{1}{105}\}$, \eqref{J5pp} describes a 9th-order pseudo-peakon.

Figures 4(a)--4(f) illustrate the $b$-independent pseudo-peakon \eqref{J5pp} with the following parameter selections:
$\{a_1 = 10, a_2 = -7, a_3 = 1\}$, $\{a_1 = 1, a_2 = -5, a_3 = 1\}$, $\{a_1 = 1, a_2 = 1, a_3 = 1\}$, $\{a_1 = \frac{1}{2}, a_2 = \frac{1}{5}, a_3 = \frac{1}{10}\}$, $\{a_1 = \frac{3}{7}, a_2 = \frac{2}{21}, a_3 = \frac{1}{105}\}$, and $\{a_1 = \frac{1}{2}, a_2 = \frac{1}{5}, a_3 = -2\}$, respectively. Meanwhile, Figs. 4(g)--4(h) demonstrate that the 9th-order pseudo-peakon solution \eqref{J5pp} with $\{a_1 = \frac{3}{7}, a_2 = \frac{2}{21}, a_3 = \frac{1}{105}\}$ exhibits continuous derivatives of $v$ up to the eighth order.
\begin{figure}
\begin{center}
\subfigure{
\includegraphics[width=0.45\textwidth,height=3cm]{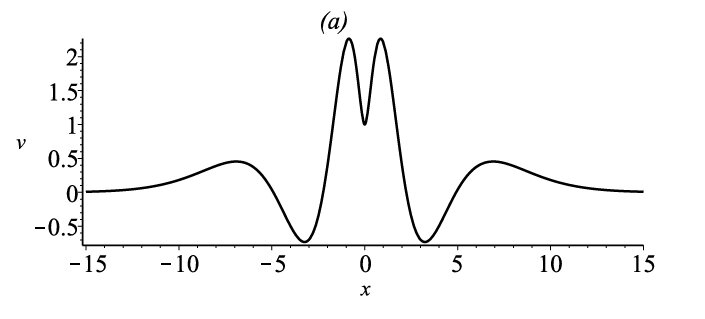}}
\centering \subfigure{
\includegraphics[width=0.45\textwidth,height=3cm]{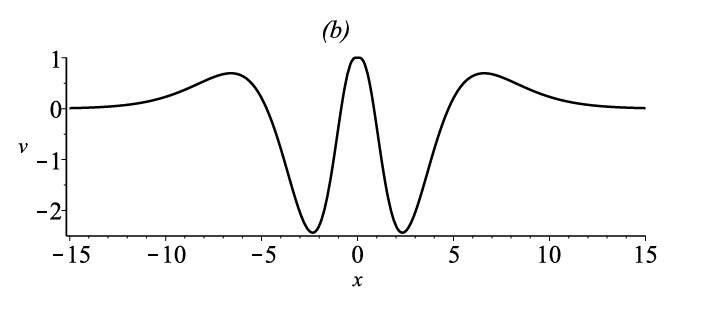}}
\subfigure{
\includegraphics[width=0.45\textwidth,height=3cm]{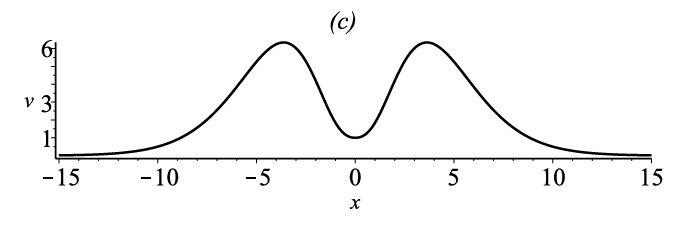}}
\centering \subfigure{
\includegraphics[width=0.45\textwidth,height=3cm]{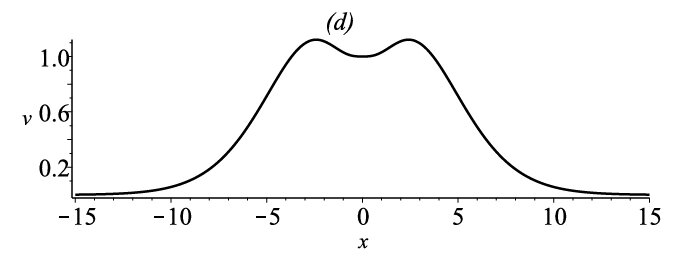}}
\subfigure{
\includegraphics[width=0.45\textwidth,height=3cm]{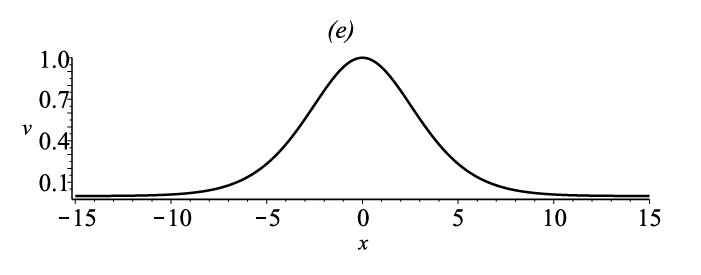}}
\centering \subfigure{
\includegraphics[width=0.45\textwidth,height=3cm]{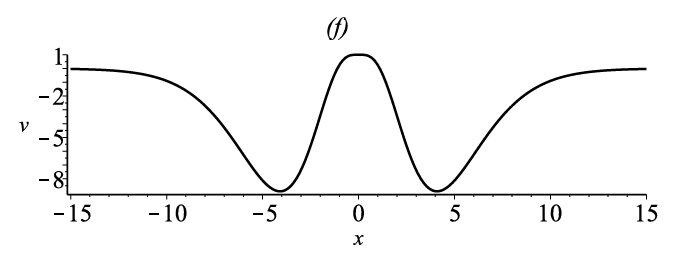}}
\centering \subfigure{
\includegraphics[width=0.45\textwidth,height=3cm]{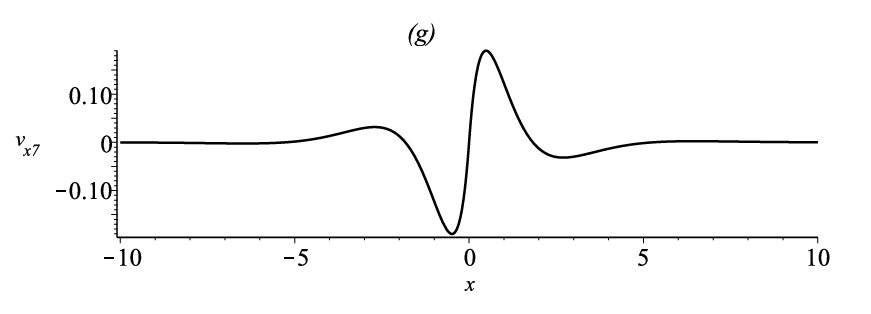}}
\centering \subfigure{
\includegraphics[width=0.45\textwidth,height=3cm]{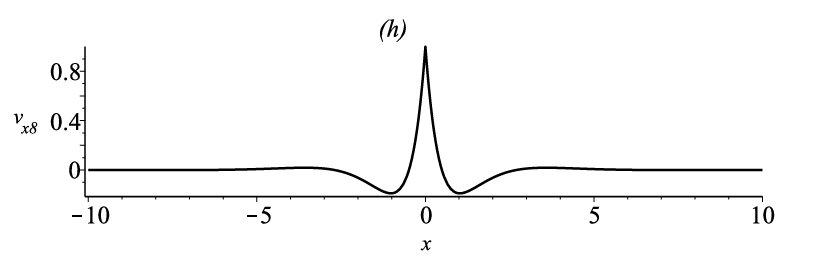}}
\caption{
    (a) Pseudo-peakon with four smooth peaks expressed by \eqref{J5pp} and $\{a_1 = 10,\ a_2 = -7,\ a_3 = 1\}$;
    (b) Pseudo-peakon with three smooth peaks given by \eqref{J5pp} and $\{a_1 = 1,\ a_2 = -5,\ a_3 = 1\}$;
    (c) M-shaped pseudo-peakon expressed by \eqref{J5pp} with $\{a_1 = 10,\ a_2 = 1,\ a_3 = 1\}$;
    (d) Pseudo-peakon with double smooth peaks expressed by \eqref{J5pp} and $\{a_1 = \frac{1}{2},\ a_2 = \frac{1}{5},\ a_3 = \frac{1}{10}\}$;
    (e) Single smooth peaked pseudo-peakon expressed by \eqref{J5pp} with $\{a_1 = \frac{3}{7},\ a_2 = \frac{2}{21},\ a_3 = \frac{1}{105}\}$;
    (f) W-shaped pseudo-peakon expressed by \eqref{J5pp} with $\{a_1 = \frac{1}{2},\ a_2 = \frac{1}{5},\ a_3 = -2\}$;
    (g) The seventh-order derivative of $v$ with respect to $x$ of (e);
    (h) The eighth-order derivative of $v$ with respect to $x$ of (e).
}
\label{fig4}
 \end{center}
\end{figure}

The second solution of \eqref{aJ5}, corresponding to \( A = 0 \), \( B \neq 0 \), and \( C = 0 \), is associated with the \( b \)-independent peakon solution of \eqref{5bF}:
\begin{align}
v &= \frac{c}{3493276440} \left( 4120035|x-ct|^4 + 65760592|x-ct|^3 + 490916844|x-ct|^2 \right. \nonumber \\
&\quad \left. + 1972076400|x-ct| + 3493276440 \right) \mathrm{e}^{-|x - ct|}. \label{5p}
\end{align}

The third solution of \eqref{aJ5}, for \( A \neq 0 \), \( B \neq 0 \), and \( C = 0 \), is related to the \( b \)-dependent peakon solution of \eqref{5bF}. For simplicity, we present only an approximate form of the \( b \)-dependent peakon solution:
\begin{equation}
v = c \left( a_0 + a_1|x - ct| + a_2|x - ct|^2 + a_3|x - ct|^3 + a_4|x - ct|^4 \right) \mathrm{e}^{-|x - ct|}, \label{5pb}
\end{equation}
where the coefficients are defined as
\begin{align*}
a_1 &\approx(2.888b^2 + 0.1728b^3 - 1.979b + 78.94) \frac{a_0^2}{10^3b^2} + (5.597b^2 + 1.06b - 1.579) \frac{a_0}{10b^2} \nonumber \\
&\quad - 0.104\frac{1}{b} + 0.07894\frac{1}{b^{2}}, \\
a_2 &\approx (3.912b^2 + 0.234b^3 - 2.681b + 106.9) \frac{a_0^2}{10^3b^2} + (1.36b^2 + 0.3439b - 2.139) \frac{a_0}{10b^2} \nonumber \\
&\quad - 0.0317\frac{1}{b} + 0.1069\frac{1}{b^2}, \\
a_3 &\approx (1.461b^2 + 0.08737b^3 - 1.001b + 39.92) \frac{a_0^2}{10^3b^2} + (0.1737b^2 + 0.02133b - 0.7984) \frac{a_0}{10b^2} \nonumber \\
&\quad - 0.001132\frac{1}{b} + 0.03992\frac{1}{b^2}, \\
a_4 &\approx (1.986b^2 + 0.1188b^3 - 1.361b + 54.29) \frac{a_0^2}{10^4b^2} + (1.007b^2 - 0.2496b - 10.86) \frac{a_0}{10^3b^2} \nonumber \\
&\quad + 0.0003858\frac{1}{b} + 0.005429\frac{1}{b^2}.
\end{align*}

The parameter $a_0$ is exactly related to $b$ through the equation
\begin{equation}
(c_1b^3 + c_2b^2 - c_3b + c_4)a_0^3 + (2c_3b - c_2b^2 - 3c_4)a_0^2 + (3c_4 - c_3b)a_0 - c_4 = 0, \label{ra0}
\end{equation}
where the constants $\{c_1,\ c_2,\ c_3,\ c_4\}$ are fixed by
\begin{align*}
c_1 &= 614878544608436385388895742565764444000, \\
c_2 &= 10279470885893649824891268018213135542160, \\
c_3 &= 7045277220996996124849013446715176502633, \\
c_4 &= 280965775971808797424508037702679295549980.
\end{align*}

It can be proven that for every real value of \( b \), there exist one real and two conjugate complex solutions \( a_0 \) of \eqref{ra0}. This procedure can be extended to larger values of \( J \). Through extensive calculations, we find that for every odd integer \( J = 2n + 1 \) with \( n \geq 1 \), in addition to the \( b \)-independent pseudo-peakon solution \eqref{Conj}, there exists one real \( b \)-independent peakon, one real \( b \)-dependent peakon and $J-3=2(n-1)$ complex \(b\)-dependent peakons. Similarly, for every even integer \( J = 2n + 2 \) with \( n \geq 1 \), in addition to the \( b \)-independent pseudo-peakon solution \eqref{Conj}, there exist one real \( b \)-independent peakon, two real \( b \)-dependent peakons and $J-4=2(n-1)$ complex \(b\)-dependent peakons. These conclusions have been verified for \( J \leq 14 \) using the computer algebra software MAPLE.
\section{Summary and discussions}
In summary, we have explored the rich structures of peakon and pseudo-peakon solutions for a class of higher-order \(b\)-family equations \eqref{JbF}, referred to as the \(J\)-th \(b\)-family (\(J\)-bF) equations. Our primary focus has been on the conjectures 1--3 that the \(J\)-bF equation admits one \(b\)-independent pseudo-peakon solution of the form \eqref{Conj}, one \(b\)-independent peakon solution of the form \eqref{Conj2} and one \(b\)-dependent peakon solution of the form \eqref{Conj3}. These conjectures have been analytically verified for \(J \leq 14\) and/or \(J \leq 9\) using the computer algebra software MAPLE, and it generalizes previous results for the cases of \(J = 1\) and \(J = 2\). While a rigorous proof for arbitrary \(J\) remains an open problem, the results presented in this paper strongly support the validity of the conjectures.

The $b$-independent pseudo-peakon solution \eqref{Conj} is a 3rd-order pseudo-peakon for general arbitrary constants $a_i$, except for some special cases. When the first two constants $a_1$ and $a_2$ satisfy the constraint \eqref{a12}, the solution \eqref{Conj} becomes a 5th-order pseudo-peakon solution. The 7th-order pseudo-peakon solutions are contained in \eqref{Conj} when the first four constants $a_i$ ($i=1, 2, 3, 4$) satisfy two constraints \eqref{a12} and \eqref{a34}. Furthermore, $(2n+1)$th-order pseudo-peakons can be found by introducing $n-1$ constraints:

\[
\left.\frac{\partial^{2i-1} v}{\partial x^{2i-1}}\right|_{\xi\rightarrow 0^+} -
\left.\frac{\partial^{2i-1} v}{\partial x^{2i-1}}\right|_{\xi\rightarrow 0^-} = 0, \quad i=2, \ldots, n \geq 2.
\]

Beyond the pseudo-peakon solutions, we have also identified the existence of both \(b\)-independent and \(b\)-dependent peakon solutions. The \(b\)-independent peakons are characterized by their independence from the parameter \(b\), and their explicit forms have been derived for various values of \(J\). These solutions exhibit non-smooth peaks and are distinct from the pseudo-peakons, which possess higher-order derivative discontinuities.

In addition to the \(b\)-independent peakons, we have discovered \(b\)-dependent peakon solutions, whose forms explicitly depend on the parameter \(b\). Through detailed and intricate calculations, we have shown that for odd integers \(J\) and any real \(b\), there exists only one real \(b\)-dependent peakon solution, whereas for even integer cases, there are two real \(b\)-dependent peakon solutions. This distinction highlights the nuanced relationship between the parameters \(b\) and \(J\) and the structure of the peakon solutions.

The existence of both \(b\)-independent and \(b\)-dependent peakons, along with the pseudo-peakon solutions, underscores the rich dynamical structure of the \(J\)-bF equations. These solutions not only generalize previous results for lower-order equations but also provide new insights into the interplay between nonlinearity and dispersion in higher-order wave models.

Future research directions in the study of peakon systems related to this paper could focus on the following areas:

   \textit{Higher-order generalizations:} Investigate all possible higher-order extensions of all known peakon systems and establish the corresponding conjectures similar to \eqref{Conj}, \eqref{Conj2} and \eqref{Conj3} for their possible peakon or pseudo-peakon solutions for all higher extensions.

   \textit{Proof of conjectures:} Rigorously prove the conjectures related to these generalized peakon systems. This would involve advanced analytical techniques to validate theoretical predictions.

  \textit{Interactions of peakons and pseudo-peakons:} Study the possible interactions between different types of peakons and pseudo-peakons. Understanding these interactions could reveal new dynamics and stability properties.

  \textit{Stability of new peakons and pseudo-peakons:} Analyze the stability of newly discovered peakon and pseudo-peakon solutions. This includes both linear and nonlinear stability analysis to determine their robustness under different types of perturbations.

 \textit{Integrable and/or non-integrable properties:} Explore the integrable and non-integrable properties of various generalized models and their corresponding multi-peakon dynamical systems.

 \textit{Mathematical structures and geometric properties:} Investigate the underlying mathematical structures, including geometric properties, of the generalized models. This could involve studying the Hamiltonian structures, conservation laws, symmetries, symplectic geometry, and other geometric aspects.

 \textit{Physical applications:} Explore potential physical applications of these generalized peakon models. This could include applications in fluid dynamics, nonlinear optics, and other areas where solitary waves, peakons, and pseudo-peakons play a crucial role.

These research directions would significantly advance our understanding of peakon systems and their broader implications in both mathematics and physics.

\begin{acknowledgments}
The work was sponsored by the National Natural Science Foundations of China (Nos.12235007, 12271324, 11975131). The authors are indebt to thank Profs. Z. J. Qiao, Q. P. Liu, B. F. Feng, X. B. Hu, M. Jia, H. L. Hu and X. Z. Hao for their helpful discussions.
\end{acknowledgments}


\end{document}